\documentclass[a4paper,12pt]{article}
\usepackage[utf8]{in
    putenc}
\usepackage{cancel}
\usepackage{ulem}
\usepackage{amsfonts}
\usepackage{amssymb}
\usepackage{graphicx}
\usepackage{amsmath,bm}
\usepackage{enumerate}
\usepackage{mathtools}
\usepackage{tikz} \usetikzlibrary{calc}
\usepackage{subfig}
\usepackage{xcolor}
\usepackage{float}
\usepackage{color}
\usepackage{here}
\usepackage{cite}

\usepackage{mathrsfs}
\usepackage{float,epsfig}
\usepackage{dcolumn}% Align table columns on decimal point
\usepackage{pgfplots}
\usepackage{graphicx}% Include figure files
\usepackage{bm}% bold math
\usepackage{amsmath,amssymb,amsthm}
\usepackage[colorlinks=true,linkcolor=blue,citecolor=red]{hyperref}
\newcommand{\be}{\begin{equation}}
\newcommand{\ee}{\end{equation}}
\newcommand{\bea}{\setlength\arraycolsep{2pt} \begin{eqnarray}}
\newcommand{\eea}{\end{eqnarray}}

\def\0{{\sst{(0)}}}
\def\1{{\sst{(1)}}}
\def\2{{\sst{(2)}}}
\def\3{{\sst{(3)}}}
\def\4{{\sst{(4)}}}
\def\5{{\sst{(5)}}}
\def\6{{\sst{(6)}}}
\def\7{{\sst{(7)}}}
\def\8{{\sst{(8)}}}
\def\sst#1{{\scriptscriptstyle #1}}

\definecolor{lime}{HTML}{A6CE39}
\newcommand{\orcidicon}{%
    \begin{tikzpicture}
    \draw[lime, fill=lime] (0,0)
        circle [radius=0.16]
        node[white] {{\fontfamily{qag}\selectfont \tiny ID}};
    \draw[white, fill=white] (-0.0625,0.095)
        circle [radius=0.007];
    \end{tikzpicture}   \hspace{-2mm}
}
\newcommand\orcidAdil{{\href{https://orcid.org/0000-0001-7623-5541}{\orcidicon}}}
\newcommand\orcidHasan{{\href{https://orcid.org/0000-0001-7408-0910}{\orcidicon}}}

\makeatletter \@addtoreset{equation}{section}

\usepackage{multirow}
\begin{document}

\title{\bf \Large
  Cosmological Constant Effect on  Charged and Rotating Black Hole  Shadows }
\author{   \small A. Belhaj\orcidAdil\!\! $^{1}$\footnote{a-belhaj@um5r.ac.ma},  M. Benali\orcidAdil\!\!$^{1}$\footnote{mohamed\_benali4@um5.ac.ma},  A. El Balali$^{1}$\footnote{anaselbalali@gmail.com},  W. El Hadri$^{1}$\footnote{elhadriwijdane4@gmail.com},  H. El Moumni\orcidHasan\!\!$^{2}$\thanks{h.elmoumni@uiz.ac.ma}\footnote{ Authors in alphabetical order.}
	\hspace*{-8pt} \\
	%EndAName
	{\small $^1$ D\'{e}partement de Physique, Equipe des Sciences de la mati\`ere et du rayonnement,
		ESMaR}\\ {\small   Facult\'e des Sciences, Universit\'e Mohammed V de Rabat, Rabat,  Morocco} \\
	{\small $^{2}$  EPTHE, Physics Department, Faculty of Science,  Ibn Zohr University, Agadir, Morocco} } \maketitle

	\begin{abstract}
		{\noindent}
		Motivated by  recent  astrophysical observations, we    investigate the shadow behaviors   of  four dimensional charged rotating  black holes  with a  cosmological constant. This study is made    in terms  of a reduced  moduli space parameterized by the
charge  and the rotation parameters.  For fixed observers, we   analyse in  some details   the   shadow behaviors   and   the corresponding naked singularities  of    Kerr-Newman       and  Kerr-Sen four-dimensional   black holes   in  Anti de Sitter backgrounds.  Then,   a  comparative  discussion is   provided    by   computing the   geometrical observables  and the  energy emission rate.
	\end{abstract}
%\newpage	
	\tableofcontents
%\newpage
\section{Introduction}
Four dimensional   black holes   have received an increasing interest supported not only  by many physical models including superstrings and related theories \cite{1,2}, but also by  the international Event Horizon Telescope
(EHT) collaboration   having  unveiled the first shadow image of
a supermassive black hole, located  in the center of galaxy M87 \cite{3,4}.

 The strong evidence of the existence of supermassive black holes at the center of most galaxies in the universe, including the Milky Way \cite{milkyway}, has supported   several  investigations  in different models of gravity \cite{Awad:2017sau, Hanafy:2015yya, Nashed:2004pn,Nashed:2003ee}. This includes  braneworld black hole \cite{Eiroa:2017uuq}, Kerr-Newmann-NUT one \cite{D1},  models with the cosmological constant where the usual formulae of  geodesics, shadows, time of time delay and  deflection angle. \cite{Bhamidipati:2018yqy, reff1,reff2,reff3}. Considering the fact that  universe  is filled or dominated by a cosmological constant, the usual expression for gravitational lensing in Friedmann-Robertson-Walker  geometries   should be  modified. In particular,  the associated expressions need to be reanalyzed  in order to  cheek the cosmological observations \cite{xx1,xx2,xx3}. On the other hand,
a particular emphasis has been put on the interplay between  such  a  physics on Anti-de-Sitter (AdS) geometries with a negative cosmological  and thermodynamics. This has opened interesting windows to develop and elaborate many links  with critical  behaviors appearing in the  black hole physics. Concretely, phase transitions of various AdS black holes have been extensively investigated showing non-trivial results. In this context, many thermodynamical quantities have been computed in order to unveil the thermodynamical properties of black holes obtained from different   gravity theories \cite{5,6}. Four and higher dimensional concrete  solutions have been dealt with  by  interpreting  the cosmological constant as  the  pressure and its conjugate as the
volume \cite{7,8}. Thermodynamics of   charged and rotating AdS black
hole systems,  controlled generally   by the mass, the charge and the angular momentum,  have been approached  using the physics of Van der Waals fluids.   Superstring models and M-theory have been also exploited  to  investigate  such properties by implementing other stringy  fields including tensor and scalar fields\cite{9,10,100}.\\
More recently, optical properties of four-dimensional black holes   have been largely  studied in connections with non-trivial backgrounds  completing the thermodynamical investigations \cite{ ref1,ref2,ref3,ref4,11,13,z0,z1}.  Precisely,  the shadows  have been considered as  a physical reality    supported by EHT  collaborations. For such reasons, shadow and deflection angle behaviors  of various  charged and rotating black holes have been discussed using different methods.  In particular, it has been revealed that  the shadows of non-rotating black holes involve a circular geometry. However, such a geometry can be deformed by introducing  rotation parameters needed for engineering  spinning solutions.
 In this models, the size of such a geometry depends  also on certain parameters associated with  external sources including dark energy (DE) and dark matter (DM) \cite{DE,DM}.
  A close inspection shows   that  the  charged  rotating  AdS black hole  shadows,   in light of   observations,  could   provide insight to  the spacetime structure  and information  on the corresponding physics. In particular, the geodesics can be linked  to two-point correlations in AdS/CFT context\cite{Balasubramanian:1999zv}. Therefore,   the analysis of  such  shadows  could  be considered  as a  useful tool to  not only explore the   astrophysical black holes but also  to compare  alternative theories with general relativity.  In  \cite{reff1,reff2,reff3}, it has been  focussed  on angular radius of spherical black hole shadows with respect to   comoving observers.

The aim of this work is to  investigate  the shadow   behaviors of four-dimensional  charged rotating  black holes with a cosmological constant  in terms of a reduced moduli space parameterized by the
charge  and the rotation parameters.  Inspired by the recent work on the  AdS backgrounds \cite{14},  we   first     elaborate,  in some details,    explicit models treating  the  Kerr-Newman (KN)  solution  and the Kerr-Sen (KS)  black hole and the associated naked singularity shadow.    Then,    we discuss and   compare the   geometrical observables  and the  energy emission rate of both black holes.      In  this work, we use
dimensionless units  ($G = \hbar =c = 1$)  and  the recent method   dealing with  celestial coordinates for a  cosmological constant  backgrounds \cite{D1,D2,D3}.

 The paper is structured  as follows. In section 2,  we  investigate  the shadow   behaviors of  KN-AdS  and KS-AdS   black holes with a negative cosmological constant, respectively.    Then,  we discuss the naked singularity and its related shadow picture.  In section 3, we analyse geometrical observable and  the corresponding  energy emission rate.  In section 4, we  inspect the cosmological  constant effect on     optical  aspects  for  (A)dS backgrounds. The last section is devoted to  conclusions and open questions
\section{Shadow behaviors of charged and  rotating black holes}
In this section, we study the optical aspects of charged and rotating black holes in four dimensions.  Concretely, we discuss the shadow geometrical pictures in terms of many parameters including the cosmological constant.
\subsection{Shadows of Kerr-Newman AdS black hole}\label{secKN}
We start by considering the photon geodesics around the   four-dimensional      KN  black hole with a   cosmological constant,
being  a charged generalization of the Kerr  black model.  To get such a solution,  one should exploit   the Einstein-Maxwell modified action  which reads as
\begin{equation}
\mathcal{I}=-\frac{1}{16\pi G}\int_{M}dx^{4}\sqrt{-g}\left[ R-F^{2}-2\Lambda %
\right],   \label{action}
\end{equation}
where    $F=dA$ denotes  the field strength of    the  gauge potential
1-form.  $ \Lambda$  is   the cosmological constant.  It is worth noting that  two   solutions  can arise depending on  $ \Lambda$.   For $ \Lambda> 0$,  the solution will be called Kerr-Newman de Sitter (KN-dS). However,   $ \Lambda <0$ generates a solution   referred to as   Kerr-Newman Anti  de Sitter (KN-AdS) which will be  investigated in certain  details through this work.  As usually,    the  variation of  the above   action with respect to the
metric tensor $g_{\mu\nu}$     can give   KN solutions. According to  \cite{17,18},
 the Boyer-Lindquist coordinates   provide the following line element
\begin{eqnarray}
\label{ds}\nonumber
ds^{2}&=&-\frac{\Delta_{r}}{\Sigma} \left(dt - \dfrac{a}{\Xi} \sin^2\theta d\phi \right)^2 +\Sigma  \left(\frac{dr^2}{\Delta_{r}}+\frac{d\theta^2}{\Delta_\theta} \right) + \frac{\Delta_\theta \sin^2\theta}{\Sigma} \left( a dt -\frac{(r^2+a^2)}{\Xi}d\phi \right)^2,
\end{eqnarray}
where  one has $
\Sigma=r^2+a^2\cos^2\theta$ and $\Xi=1+\frac{\Lambda}{3} a^2$. However, the $\Delta$ functions are given by
\begin{equation}\label{deltazz}
\Delta_\theta=1+\frac{\Lambda}{3} a^2\cos^2\theta,\quad\quad
\Delta_r=(r^2+a^2)(1-\frac{\Lambda r^2}{3}) -2mr+Q^2.
\end{equation}
Here,  $ m$, $a$ and $Q$ are the mass parameter, the angular momentum per unit
mass and the  charge, respectively. According to shadow black hole  activities, the photon equation of motion  on such   a background can be  elaborated   using the   Hamilton-Jacobi equation
\begin{equation}
\label{ks1}
\frac{\partial S}{\partial \tau}+\frac{1}{2}g^{\mu\nu}\frac{\partial S}{\partial x^\mu}\frac{\partial S}{\partial x^\nu}=0,
\end{equation}
where $\tau$ is the affine parameter associated with the geodesics. $S$  being known by  the Jacobi action is given by
\begin{equation}
\label{ks2}
S=-Et+L\phi+S_r(r)+S_\theta(\theta),
\end{equation}
where  $E=-p_t$ and $L=p_\phi$  are  the conserved  total energy and  the  conserved angular momentum of the photon,  respectively, where $p_\mu$ is  its four-momentum.  They are geodesic
constants of motion. $S_r(r)$ and $S_\theta(\theta)$  are functions  depending  only on $r$ and $\theta$ variables, respectively.
To get the complete   relations  of null geodesics, a separation method is needed  which  could be supported by  the Carter mechanism \cite{Carter,chandrasekhar1998mathematical}.  The  illustration of the  black hole shadow  geometries requires  two dimensionless impact parameters   expressed  as
\begin{equation}
\label{xe}
 \xi_{KN}=\frac{L}{E}, \hspace{1.5cm}\eta_{KN}=\frac{\mathcal{K}}{E^2},
\end{equation}
where $\mathcal{K}$  denotes a  separable  constant
analogue  to the  Carter  one  reported  in \cite{Carter}. The subscript ${KN}$   stands for  the Kerr-Newmann black hole solution.  To obtain the corresponding relations, ceratin calculations should be performed.  Indeed, they  give the  null geodesics equations
\begin{align}
\Sigma \frac{d  \, t}{d \tau}& =  E \left[ \frac{\left(r^2 +a^2 \right)\left[ \left(r^2 +a^2 \right)- a \xi_{KN} \Xi  \right] }{\Delta_r} +  \frac{ a \left(\xi_{KN} \Xi - a  \sin^2 \theta \right) }{ \Delta_\theta} \right], \\
\Sigma  \frac{d  \, r}{d \tau} &=\sqrt{\mathcal{R}_{KN}(r}),\\
 \Sigma\frac{d  \, \theta}{d \tau} & =\sqrt{\Theta_{KN}(\theta)},\\
\Sigma \frac{d  \, \phi}{d \tau} &= E \;\Xi \left[ \frac{a(\left(r^2 +a^2 \right) -a \xi_{KN} \Xi)}{\Delta_r} + \frac{ \xi_{KN} \Xi-a \sin^2 \theta}{\sin^2 \theta \Delta_\theta}  \right].
\end{align}

 In these relations,  $\mathcal{R}_{KN}(r)$ and ${\Theta}_{KN}(\theta)$   describing the radial and  the polar motion   read as
 \begin{align}
\mathcal{R}_{KN}(r)&=E^2\left[ \left[ \left( r^2 +a^2 \right)  -a \xi_{KN} \Xi  \right]^2- \Delta_r\eta_{KN}   \right],\\
 \Theta_{KN}(\theta)&= E^2 \left[ \eta_{KN} \Delta_\theta  - \csc^2 \theta\left( a\sin^2 \theta - \xi_{KN} \Xi\right )^2 \right].
\end{align}
It is known that   the unstable circular orbit can  determine the boundary
of  the  black hole geometric shapes using   the constraints
\begin{equation}\label{xx1}
\mathcal{R}_{KN}(r)\Big|_{r=r_0}=\frac{d\mathcal{R}_{KN}(r)}{d r}\Big|_{r=r_0}=0,
\end{equation}
where $r_0$ represents the circular orbit radius of the photon\cite{chandrasekhar1998mathematical,A7,H}.  By solving Eqs.\eqref{xx1} and taking into account $ \Theta_{KN}(\theta)>0$ for $0\leqslant\theta\leqslant2\pi$ , we find
\begin{align}
& \eta_{KN} = \frac{16 r^2 \Delta_r}{\Delta _r^{\prime\,2}}\bigg\vert_{r=r_0},\\
& \xi_{KN}=\frac{\left(r^2+ a^2\right) \Delta_r^\prime-4r \Delta_r  }{a\Xi \Delta_r^\prime} \bigg\vert_{r=r_0},
\end{align}
where one has used the following derivation notation $\Delta_r^\prime=\frac{\partial \Delta_r}{\partial r}$. In the presence of cosmological constant $\Lambda$,  the position ($r_{ob}$\,,$\theta_{ob}$) of the observer in Boyer-Lindquit coordinates   should be fixed \cite{D3}, where  $r_{ob}$  is the radial coordinate and $\theta_{ob}$ is angular coordinate. Assuming that  the observer is in the domain of outer communication ($\Delta_r >0$), and  considering  the trajectories of light rays sent from position ($r_{ob}$\,,$\theta_{ob}$) to past,  we  explore the recent method reported  in Ref\cite{D1,D2}  to  define the orthogonal tetrads ($e_0$,\,$e_1$,\,$e_2$,\,$e_3$) associated with  the observer  position
\begin{eqnarray}
\label{e_0}
e_0 & = & \frac{ (r^2+a^2)\partial_t+a\Xi\partial_\phi}{\sqrt{\Delta_r \Sigma}} \bigg\vert_{(r_{ob}\,,\theta_{ob})},\\
\label{e_1}
e_1 & = & \frac{\sqrt{\Delta_\theta}}{\sqrt{\Sigma}} \partial_\theta\bigg\vert_{(r_{ob}\,,\theta_{ob})},\\
\label{e_2}
e_2 & = &  -\frac{ a\sin^2{\theta}\partial_t+\Xi\partial_\phi}{\sqrt{\Delta_r \Sigma}\sin{\theta}} \bigg\vert_{(r_{ob}\,,\theta_{ob})}, \\
\label{e_3}
e_3 & = &-\frac{\sqrt{\Delta_r}}{\sqrt{\Sigma}} \partial_r\bigg\vert_{(r_{ob}\,,\theta_{ob})}.
\end{eqnarray}

\begin{figure}[H]
		\begin{center}
		\centering
			\begin{tabbing}
			\centering
			\hspace{8.6cm}\=\kill
			\includegraphics[scale=.5]{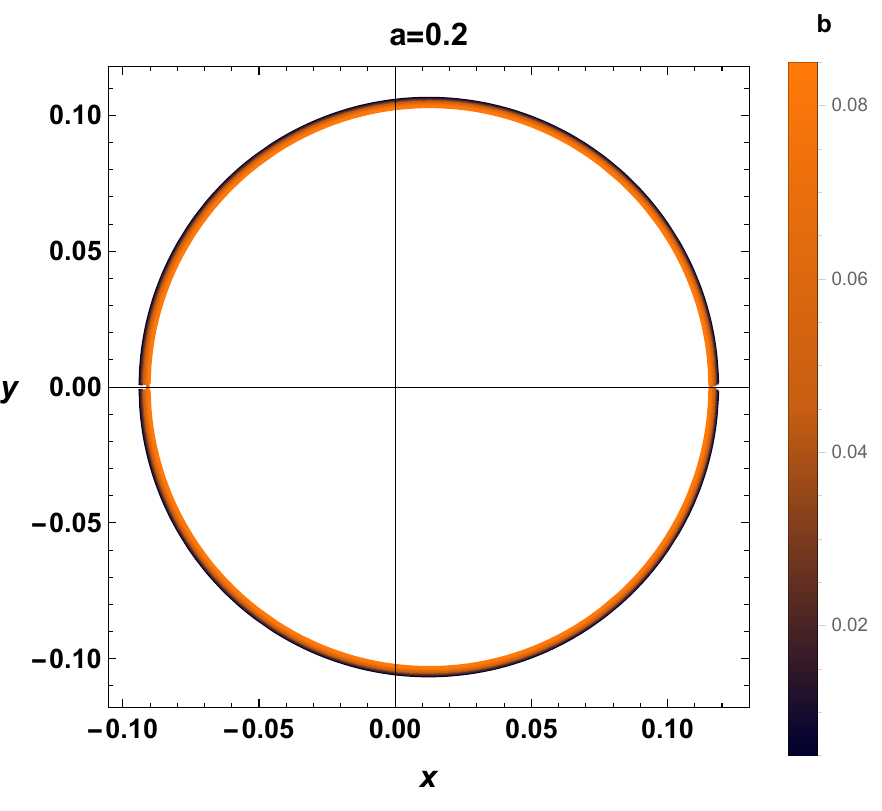} \>
			\includegraphics[scale=.5]{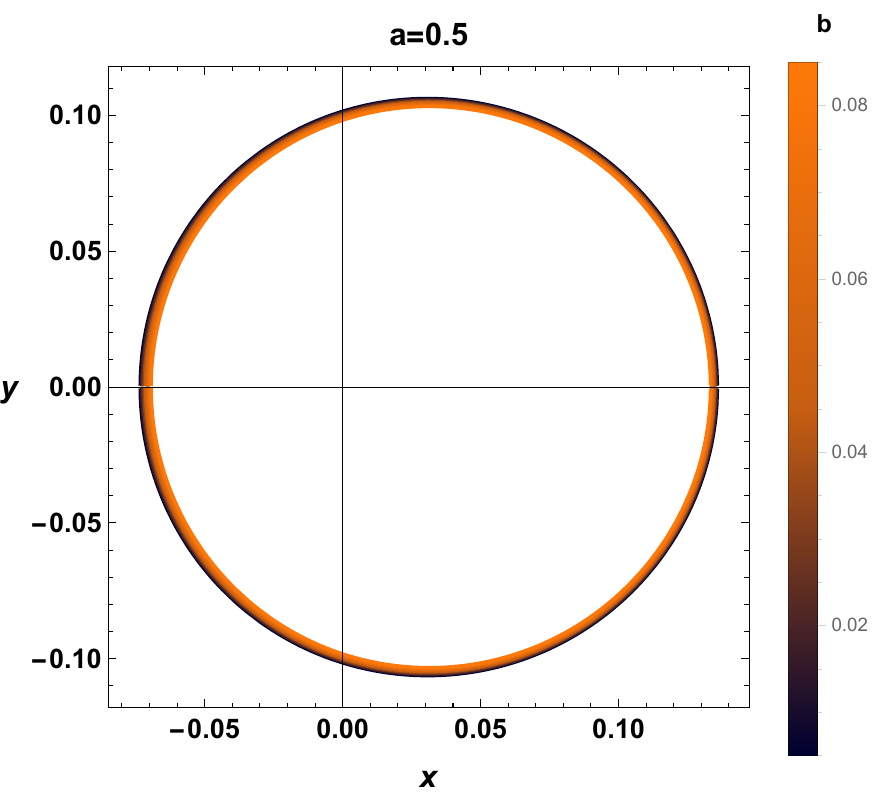} \\
			\includegraphics[scale=.5]{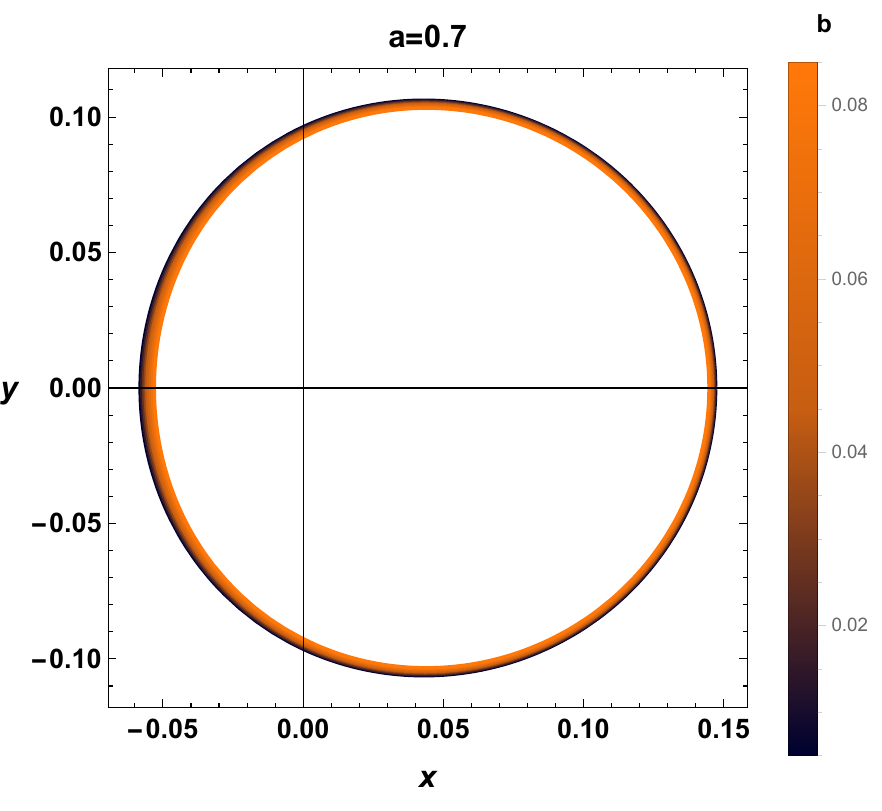} \>
			\includegraphics[scale=.5]{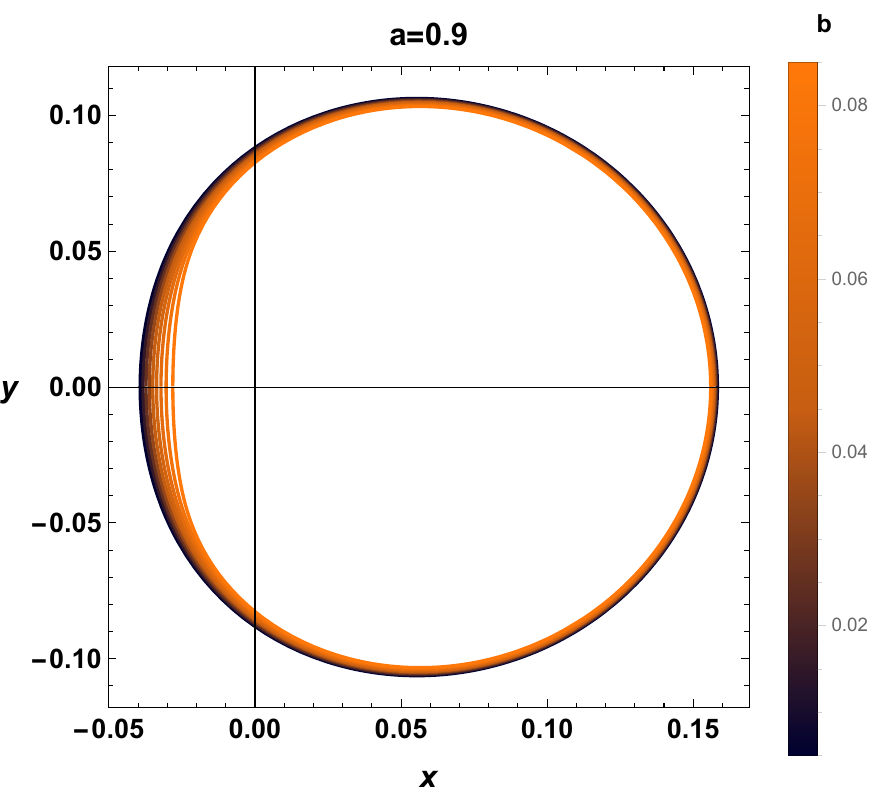} \\
		   \end{tabbing}
\caption{{\it \footnotesize Shadow behaviors of KN-AdS black holes for different values of  $a$ and  $b$ by taking $\Lambda=-10^{-4}$ and  $m=1$. The observer is positioned at $r_{ob}=50$ and $\theta_{ob}=\frac{\pi}{2}$}.}
\label{shfa1}
\end{center}
\end{figure}

In these relations,  the timelike vector $e_0$ indicates   the four-velocity of the  observer. $e_3$ denotes   the vector along the spatial direction pointing toward the center of the black hole. However,  $e_0\pm e_3$ are tangent to the direction of principal null congruences. The light ray  can be expressed  by the following   parametrization
\begin{equation}
\label{lambda}
\lambda(s)=(r(s),\theta(s),\phi(s),t(s)).
\end{equation}
In the way,   the vector tangent to $\lambda(s)$, denoted by $\dot{\lambda}$,  is  given by
 \begin{equation}
\label{lambdat}
\dot{\lambda}=\dot{r}\partial_r+\dot{\theta}\partial_\theta+\dot{\phi}\partial_\phi+\dot{t}\partial_t.
\end{equation}
Using  the celestial coordinates  $\rho$ and $\delta$ as in \cite{D1}, and the above basis vectors,  $\dot{\lambda}$ can be written
\begin{equation}
\label{lambdatt}
\dot{\lambda}=\alpha(-e_0+\sin\rho\cos\delta\,e_1+\sin\rho\sin\delta\,e_2+\cos\rho\,e_3),
\end{equation}
where $\alpha$  is a scalar factor. Combining  Eq.(\ref{lambdat}) and Eq.(\ref{lambdatt}), one gets
\begin{equation}
\label{alpha}
\alpha=g(\dot{\lambda},e_0)=\frac{1}{\sqrt{\Delta_r\Sigma}}(aL\Xi-(r^2+a^2)E) \bigg\vert_{(r_{ob}\,,\theta_{ob})}.
\end{equation}
In order to find  the celestial coordinates $\rho$ and $\delta$, we  should exploit  the equation Eq.(\ref{lambdat}) and Eq.(\ref{lambdatt}). The coefficients of $\partial_r$ and $\partial_\phi$ provide
\begin{eqnarray}
\label{rho}
\sin{\rho}&=&\sqrt{1-\frac{\Sigma^2\dot{r}^2}{(E(r^2+a^2)-aL\Xi)^2}}\bigg\vert_{(r_{ob}\,,\theta_{ob})},\\
\label{psi}
\sin{\delta}&=&\frac{\sqrt{\Delta_\theta}\sin\theta}{\sqrt{\Delta_r}\sin\rho}(\frac{\Delta_r\Sigma\dot{\phi}}{E(r^2+a^2)-aL\Xi}-a\Xi)\bigg\vert_{(r_{ob}\,,\theta_{ob})}.
\end{eqnarray}
Using the above  relations  and  implementing     $\xi_{KN}$ and $\eta_{KN}$  via   the equations of motion, one gets the  celestial coordinates $\rho$ and $\delta$ in terms of the parameters $\xi_{KN}$ and $\eta_{KN}$ as
\begin{eqnarray}
\label{rho}
\sin{\rho}&=&\frac{\pm\sqrt{\Delta_r\eta_{KN}}}{((r^2+a^2)-a\xi_{KN}\Xi)}\bigg\vert_{(r_{ob}\,,\theta_{ob})},\\
\label{psi}
\sin{\delta}&=&\frac{\sqrt{\Delta_r}\sin\theta}{\sqrt{\Delta_\theta}\sin\rho}\left(\frac{\Xi(a-\Xi\csc^2{\theta}\xi_{KN})}{a\Xi\xi_{KN}-(r^2+a^2)}\right)\bigg\vert_{(r_{ob}\,,\theta_{ob})}.
\end{eqnarray}
An examination shows that the  boundary of shadows of  such black  holes depends on many parameters including cosmological constant numerical  values.  For simplicity reasons,  we first  consider  the  AdS backgrounds   for which   $\Lambda$ is linked to the    characteristic length scale  of the AdS geometry  via $\Lambda=-\frac{3}{\ell_{AdS}^2} $. For later use,  we introduce a  twist charge parameter $b=Q^2/2m$ in the associated $\Delta_r$ function.
In this way, the shadow geometry will be controlled  by a  moduli space parameterized by $\{m,a,b\}$.  Fixing the mass, such a space reduces to $\{a,b\}$.
According to \cite{D1,D2,D3}, the boundary of   the shadow  can be visualized  using the cartesian coordinate system.
\begin{eqnarray}
\label{x}
x & = & -2 \tan{(\frac{\rho}{2})}\sin{\delta},\\
\label{y}
y & = & -2 \tan{(\frac{\rho}{2})}\cos{\delta}.
\end{eqnarray}
 In Fig.(\ref{shfa1}), the associated  shadow contours are plotted  in such a plane by exploiting    $x$ and $y$ expressions. In particular, we illustrate  the  shadow geometrical behaviors in terms of the $(a,b)$  reduced moduli space.
It has been observed that  for very small values of the rotating parameter $a$ the shadow shape  involves a   perfect  circular geometry matching whith non axisymmetric black hole shadow configurations.
When  such a  rotation parameter becomes  relevant,  the black hole shadow is distorts  by
exhibiting a so-called D-shape form.  It has been observed that  the   size depends on the $b$ parameter. Indeed,  it  decreases  by  increasing $b$.  Fixing the rotation
parameter $a$, the shadow size is decreased by the increase of the parameter $b$.

\subsection{Shadows of Kerr-Sen AdS black hole}
Here,  we deal with the  shadow behaviors of the Kerr-Sen black hole    with a negative cosmological constant    built  recently in  \cite{14}.  We refer to this  solution as   KS-AdS.  This solution has been obtained from the original KS  model    by implementing   a nonzero negative cosmological
constant \cite{ksbh}.  It is known that  this   could be   derived  from a lower energy limit of the  heterotic superstring  theory living in ten dimension.  Concretely,
the associated   black hole metric   can  be obtained  from an  action involving stringy fields, including the dilaton-axion field,  a Maxwell field  and  the B-field.  More details on the  performed  calculations  can be found in \cite{ksbh}.  Following  \cite{14},   the  line element  the  four dimensional  KS-AdS black hole,  in the coordinates ($t$,$r$,$\theta$,$\phi$), reads as
\begin{equation}
ds^{2}=-\frac{\Delta_{r}}{\Sigma} \left(dt - \dfrac{a}{\Xi} \sin^2\theta d\phi \right)^2 +\Sigma  \left(\frac{dr^2}{\Delta_{r}}+\frac{d\theta^2}{\Delta_\theta} \right) + \frac{\Delta_\theta \sin^2\theta}{\Sigma} \left( a dt -\frac{(r^2+2br+a^2)}{\Xi}d\phi \right)^2,
\end{equation}
where the  involved   terms  are given by
\begin{eqnarray}
 \Delta_{r}&=&(1-\frac{(r^2+2br)\Lambda}{3})(r^2+2br+a^2)-2mr, \qquad   \Delta_{\theta}= 1+\frac{a^{2}\Lambda}{3}\cos^{2}\theta, \\
 \Xi &=& 1+\frac{a^{2}\Lambda}{3}, \qquad   \Sigma = r^{2}+2br+a^{2}\cos^{2}\theta.
\end{eqnarray}
It is indicated  that  $m$ is  a mass function and where  $a$ is a  rotating parameter as before. While,  the parameter $b$ given  $b=Q^2/2m$ denotes now   the dilatonic scalar charge, playing the same role as the  charge parameter of the  KN black hole solution  given in Eq.\eqref{deltazz}.  Using similar techniques presented previously,  the first-order differential   equations,  describing the photon motion in such a  background, read as
\begin{align}
\Sigma \frac{d  \, t}{d \tau}& =  E \left[ \frac{\left(r^2 + 2br+a^2 \right)\left[ \left(r^2 + 2br+a^2 \right)- a \xi_{KS} \Xi  \right] }{\Delta_r} +  \frac{ a \left(\xi_{KS} \Xi - a  \sin^2 \theta \right) }{ \Delta_\theta} \right], \\
\label{1.1}
\Sigma  \frac{d  \, r}{d \tau} &=\sqrt{\mathcal{R}_{KS}(r}),\\
 \Sigma\frac{d  \, \theta}{d \tau} & =\sqrt{\Theta_{KS}(\theta)},\\
 \label{1.2}
\Sigma \frac{d  \, \phi}{d \tau} &= E \;\Xi \left[ \frac{a(\left(r^2 + 2br+a^2 \right) -a \xi_{KS} \Xi)}{\Delta_r} + \frac{ \xi_{KS} \Xi-a \sin^2 \theta}{\sin^2 \theta \Delta_\theta}  \right].
\end{align}

\begin{figure}[!htb]
		\begin{center}
		\centering
			\begin{tabbing}
			\centering
			\hspace{8.6cm}\=\kill
			\includegraphics[scale=.5]{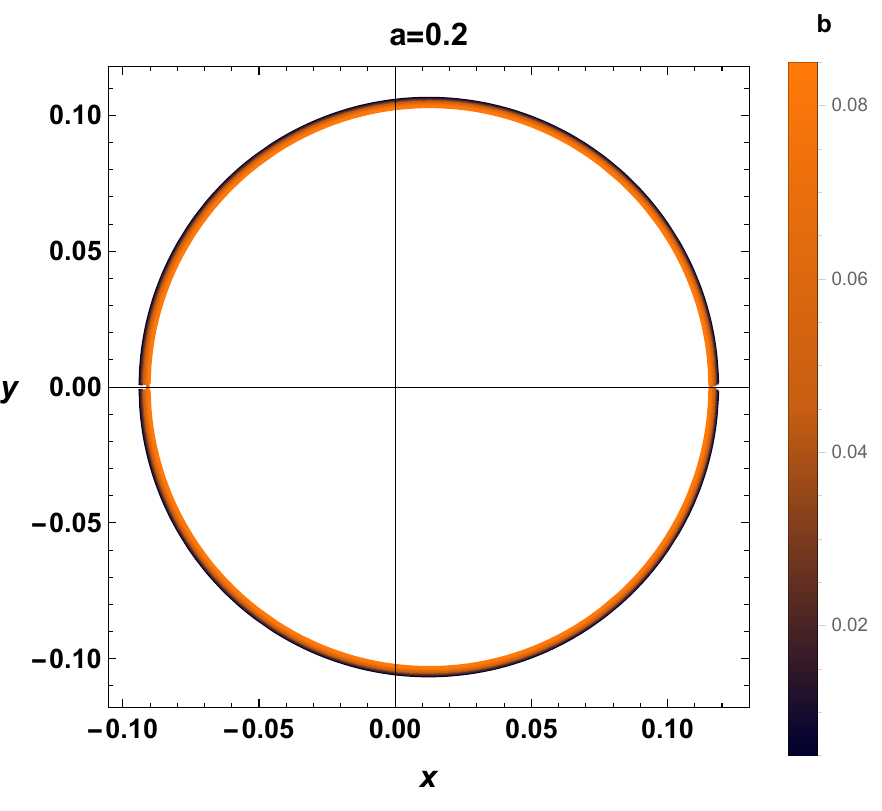} \>
			\includegraphics[scale=.5]{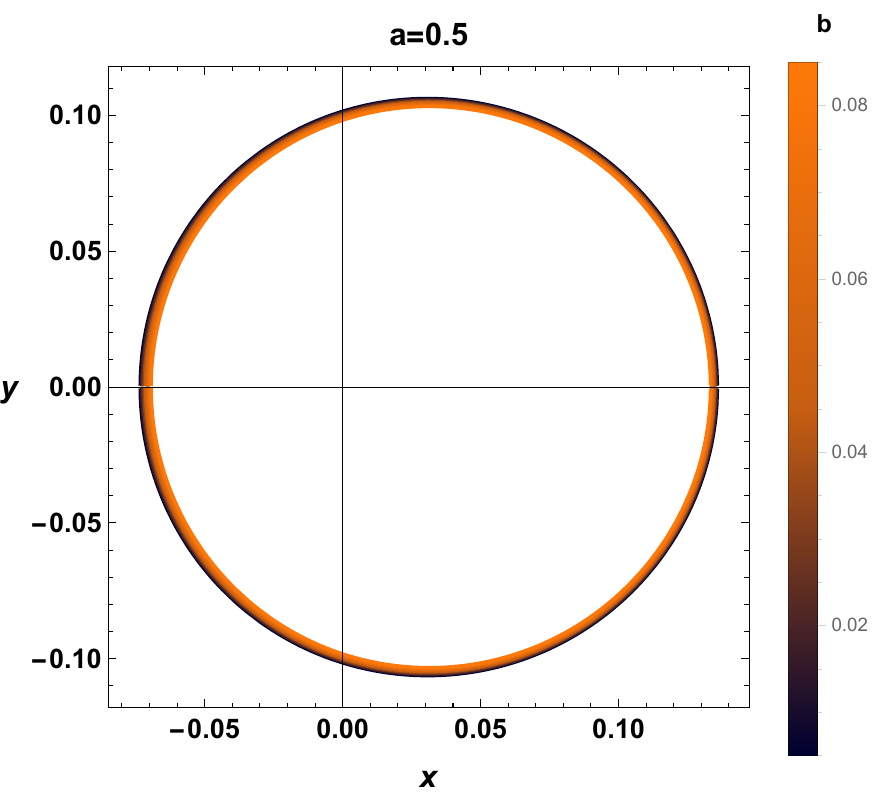} \\
			\includegraphics[scale=.5]{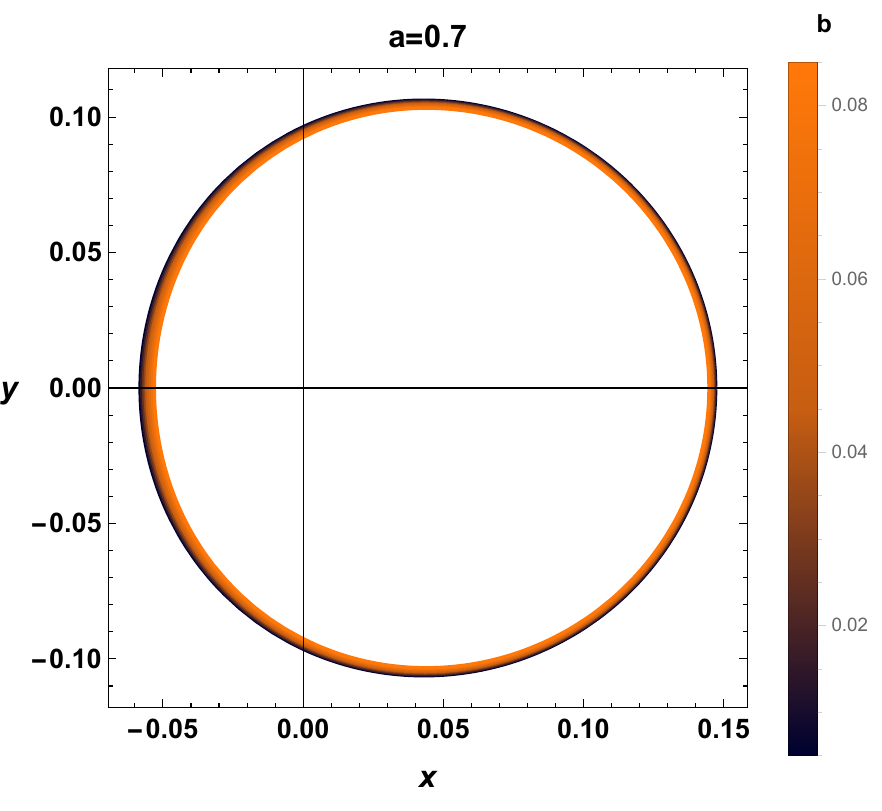} \>
			\includegraphics[scale=.5]{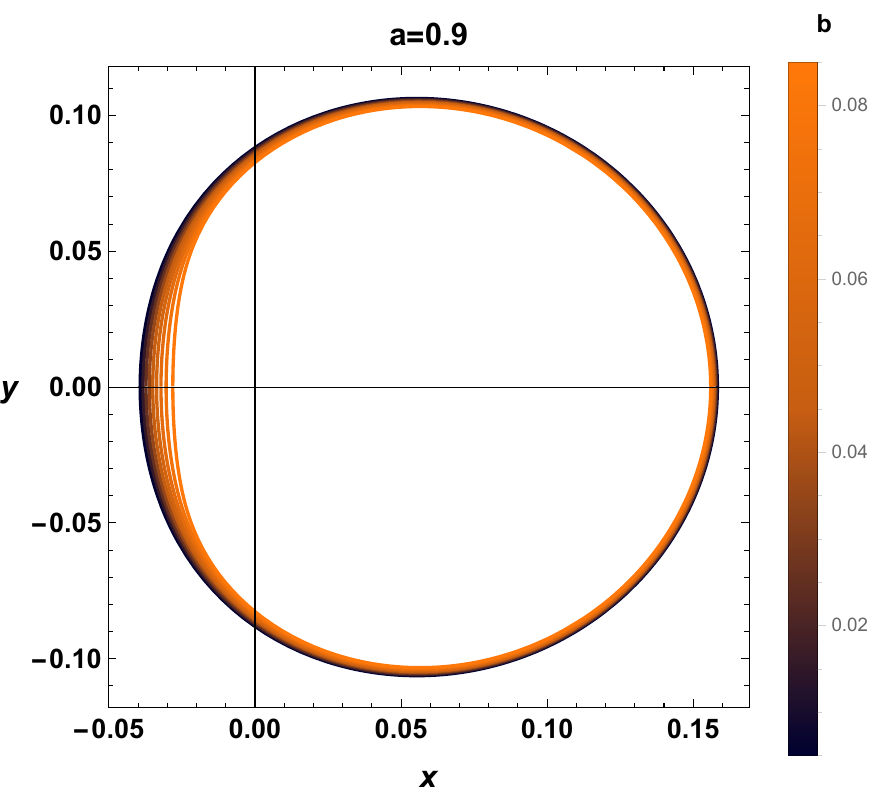} \\
		   \end{tabbing}
\caption{{\it \footnotesize Shadow behaviors  of KS-AdS black holes for different values of  $a$ and  $b$ by taking $\Lambda=-10^{-4}$ and  $m=1$. The observer is positioned at $r_{ob}=50$ and $\theta_{ob}=\frac{\pi}{2}$}}
\label{shfa}
\end{center}
\end{figure}

 In these  equations,  $\mathcal{R}_{KS}(r)$ and ${\Theta}_{KS}(\theta)$   describing  the radial and  the polar motion   take the following form
  \begin{align}
\mathcal{R}_{KS}(r)&=E^2\left[ \left[ \left( r^2 +2br+a^2 \right)  -a \xi_{KS} \Xi  \right]^2- \Delta_r\eta_{KS}   \right],\\
 \Theta_{KS}(\theta)&= E^2 \left[ \eta_{KS} \Delta_\theta  - \csc^2 \theta\left( a\sin^2 \theta - \xi_{KS} \Xi\right )^2 \right].
\end{align}
Imposing the  constraints $
\mathcal{R}_{KS}(r)\Big|_{r=r_0}=\frac{d\mathcal{R}_{KS}(r)}{d r}\Big|_{r=r_0}=0$ with $ \Theta_{KS}(\theta)>0$, the impact parameters  $\xi_{KS}$ and $\eta_{KS}$ of the KS-AdS  black hole  can be obtained. Indeed, they are given by
\begin{eqnarray}
\xi_{KS}& = & \frac{(r^2+2br+a^2)\Delta_r^\prime-4(r+b)\Delta_r} {a\Xi\Delta_r^\prime}\bigg\vert_{r=r_0},\\
\eta_{KS} & = & \frac{16(b+r)^2\Delta_r}{{\Delta_r^\prime}^2}\bigg\vert_{r=r_0}.
\end{eqnarray}
Using  similar computations, we obtain  the scalar  factor for such a black hole
\begin{equation}
\label{alpha}
\alpha=g(\dot{\lambda},e_0)=\frac{1}{\sqrt{\Delta_r\Sigma}}(aL\Xi-(r^2+2br+a^2)E) \bigg\vert_{(r_{ob}\,,\theta_{ob})}.
\end{equation}
In this stringy solution,  the celestial coordinates $\rho$ and $\delta$ are given  by

\begin{eqnarray}
\label{rho}
\sin{\rho}&=&\sqrt{1-\frac{\Sigma^2\dot{r}^2}{(E(r^2+2br+a^2)-aL\Xi)^2}}\ \bigg\vert_{(r_{ob}\,,\theta_{ob})},\\
\label{psi}
\sin{\delta}&=&\frac{\sqrt{\Delta_\theta}\sin\theta}{\sqrt{\Delta_r}\sin\rho}(\frac{\Delta_r\Sigma\dot{\phi}}{E(r^2+2br+a^2)-aL\Xi}-a\Xi)\bigg\vert_{(r_{ob}\,,\theta_{ob})}.
\end{eqnarray}
Therefore, by the help of the above equations of motion,  we can obtain the  celestial coordinates $\rho$ and $\delta$ in terms of the parameters $\xi_{KS}$ and $\eta_{KS}$. Indeed, one  finds
\begin{eqnarray}
\label{rho}
\sin{\rho}&=&\frac{\pm\sqrt{\Delta_r\eta_{KS}}}{((r^2+2br+a^2)-a\xi_{KS}\Xi)}\bigg\vert_{(r_{ob}\,,\theta_{ob})}, \\
\label{psi}
\sin{\delta}&=&\frac{\sqrt{\Delta_r}\sin\theta}{\sqrt{\Delta_\theta}\sin\rho}\left(\frac{\Xi(a-\Xi\csc^2{\theta}\xi_{KS})}{a\Xi\xi_{KS}-(r^2+2br+a^2)}\right)\bigg\vert_{(r_{ob}\,,\theta_{ob})}.
\end{eqnarray}
Taking  the limit $\ell_{AdS}$ goes to the infinity, we recover  the usual  the KS black hole equations \cite{Lan:2018lyj}. In order to visualise  the shadows of  KS-AdS  black hole, we introduce   the celestial coordinates $\rho$ and $\delta$ in the equatorial plane  as in the previous model.  The corresponding  behaviors in terms of the $(a,b)$  reduced  moduli space are plotted  in Fig.(\ref{shfa}).
It follows from such a figure  that the shadow shape is circular for slowly  rotating black hole solutions. Moreover,  its  size depends on the $b$ parameter. Indeed,  it  decreases  by  increasing $b$. It has been observed  similar D-shapes as in  the previous model for pertinent values of rotation parameter $a$.\\
Having discussed the shadow behaviors for real horizon radius values, we move to investigate  other  non-trivial configuration. It is  known that when massive matter clouds undergo a  continual gravitational collapse, the total mass collapses into a spacetime singularity.  At such a location,  the density,  pressures and spacetime curvatures become infinite \cite{singg1, singg2}. In what follows, we consider the  associated geometries.
\vspace{2 cm}

\subsection{Naked singularity shadow}

\begin{figure}[!htb]
		\begin{center}
		\centering
			\includegraphics[scale=.5]{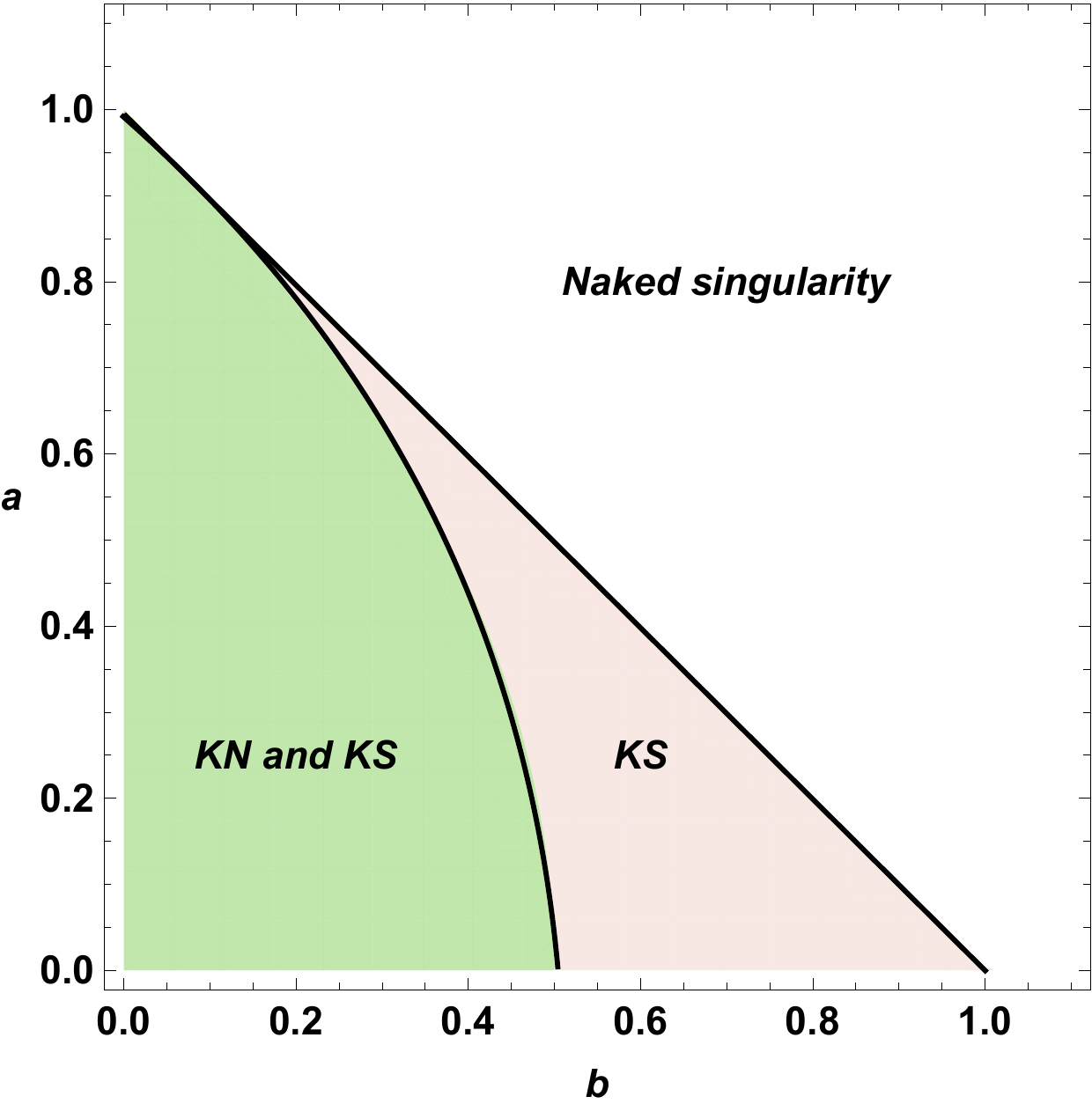}
\caption{{\it \footnotesize  {Region plot for both  KN-AdS and KS-AdS black holes as function of parameters $a$ and $b$  by taking $\Lambda=-10^{-4}$. The black  solid lines  correspond to extremal black hole cases.
}}}
\label{RN}
\end{center}
\end{figure}

Here,  we would like to  discuss the  naked singularity shadow for both  KN-AdS and KS-AdS black holes. Inspecting the shadow geometries,   the unstable spherical orbits of the photons involves  a circular geometry.   In   the naked singularity, however,   such  orbits  are represented  by  arcs, as we will see. Due  to the  horizon absence, photons being  close to both sides of the  possible  arcs can  be seen by  the observer\cite{D4,D5}. It is worth noting that  the naked singularity  appears  when the  largest root of  $\Delta_r=0$  takes complex values.  In Fig.(\ref{RN}), we illustrate the  horizon region   and the naked singularity  for  KN-AdS and KS-AdS black holes  in terms of  the parameters $a$ and $b$. It has been observed that the horizon region of KS-AdS  is bigger than  KN-AdS. This result is also obtained for both   black holes with  vanishing cosmological constant \cite{Carlos}.  The comparison of the results  reported in \cite{Carlos} and the ones presented in  Fig.(\ref{RN}) shows that the naked singularity region in Fig.(\ref{RN}) is larger.

\begin{figure}[!htb]
		\begin{center}
		\centering
			\begin{tabbing}
			\centering
			\hspace{8.6cm}\=\kill
			\includegraphics[scale=.5]{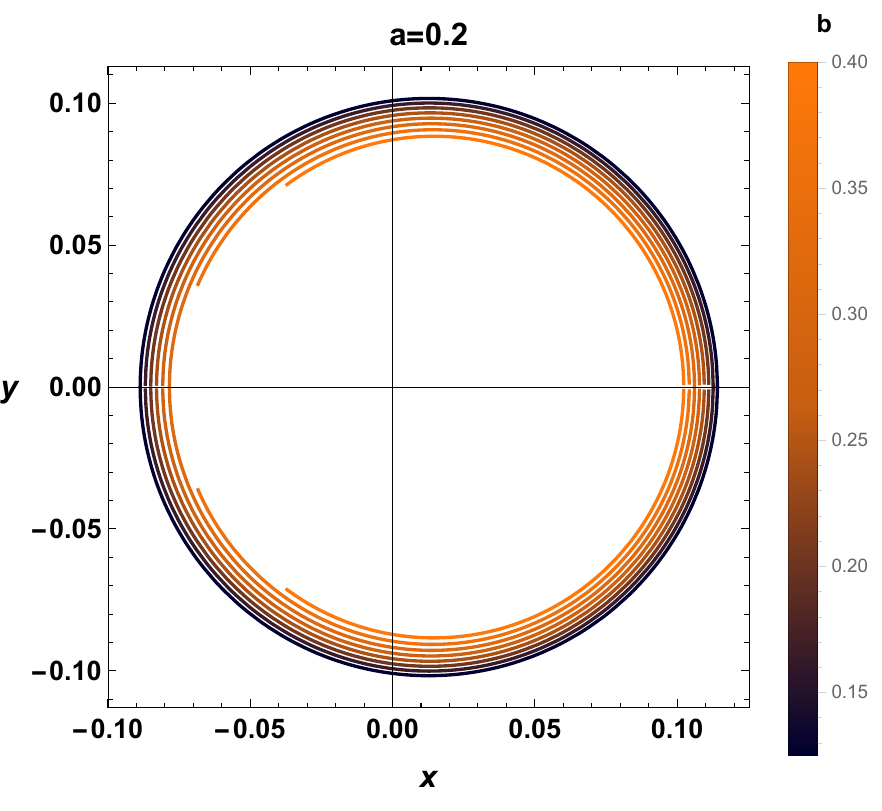} \>
			\includegraphics[scale=.5]{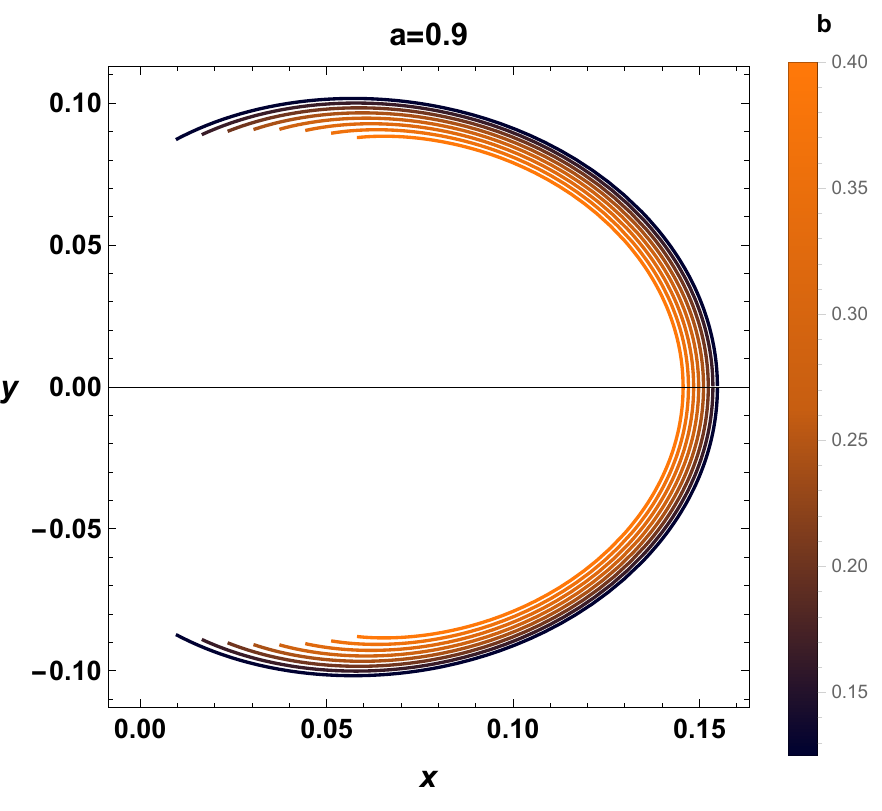} \\
			\includegraphics[scale=.5]{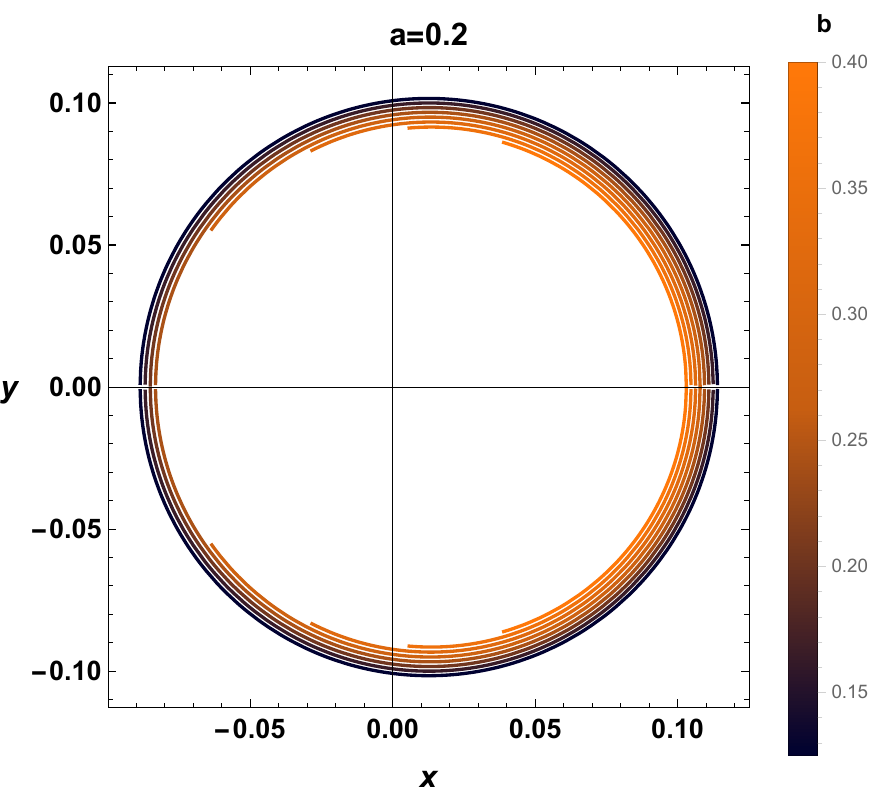} \>
			\includegraphics[scale=.5]{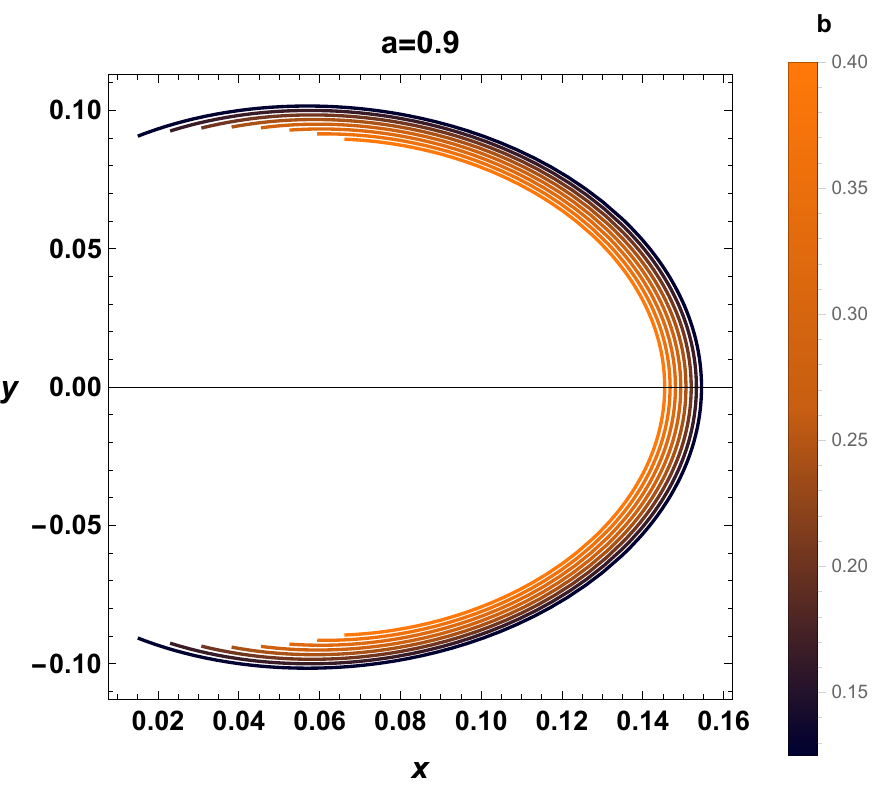} \\
		   \end{tabbing}
\caption{{\it \footnotesize Shadow behaviors  of KN-AdS (top figure) and KS-AdS (bottom figure) black holes for different values of  $a$ and  $b$ by taking $\Lambda=-10^{-4}$ and  $m=1$. The observer is positioned at $r_{ob}=50$ and $\theta_{ob}=\frac{\pi}{2}$.}}
\label{NK}
\end{center}
\end{figure}

An examination on such a solution  shows that  the $b$ parameter  is constrained by   $0.125 \leq b \leq  0.405$. Considering such a bonded limit,    the shadow as a function of the parameter $b$  and $a$  is  plotted  in Fig.(\ref{NK}). For small values of the rotating parameter, such  a singularity  appears only when  $b>0.325$ for both black holes.   Taking  $a=0.9$, however,  the naked singularity  has been observed for the above parameter range.

\vspace{3 cm}

\section{Distortion and energy emission rate: comparative study}
In this section,  we would like to provide a comparative study  associated with a negative  cosmological constant using certain regions of  the $(a,b)$  reduced  moduli space. First, we start with  the geometrical distortion  behaviors. Then, we deal with the energetic  aspects  by examining  the  energy emission rate.
\subsection{Distortion behaviors}
\begin{figure}[!ht]
		\centering
			\begin{tabbing}
			\hspace{8.8cm}\= \hspace{8.8cm}\=\kill
			\includegraphics[scale=.59]{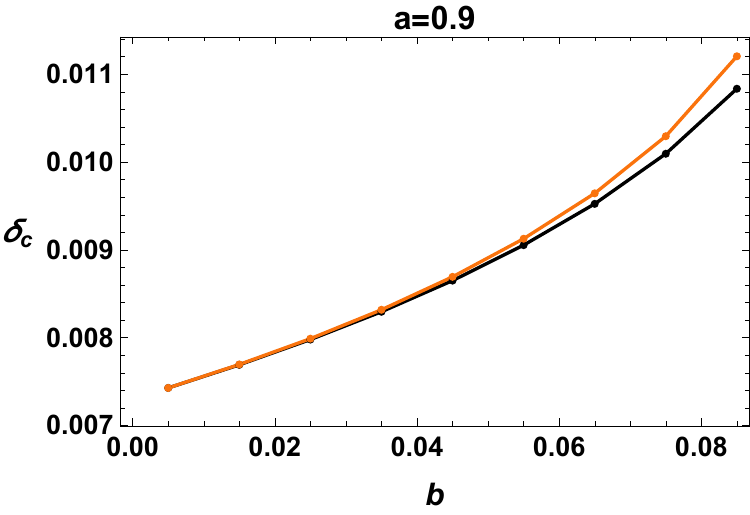} \>
			\includegraphics[scale=.59]{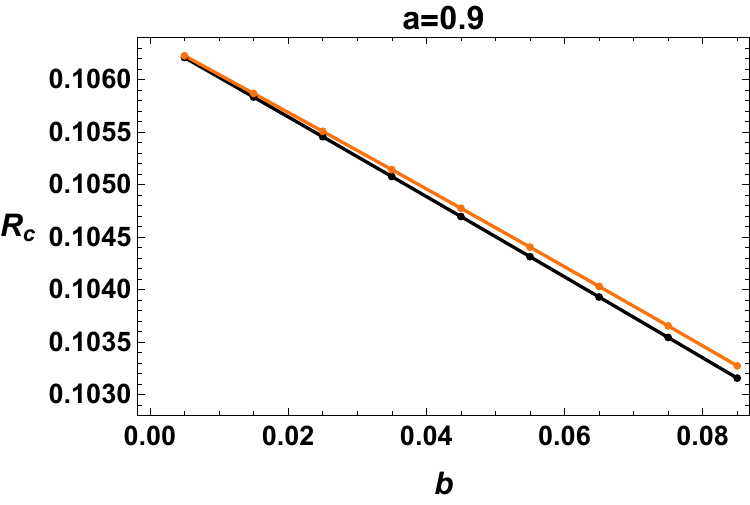} \\
			\includegraphics[scale=.59]{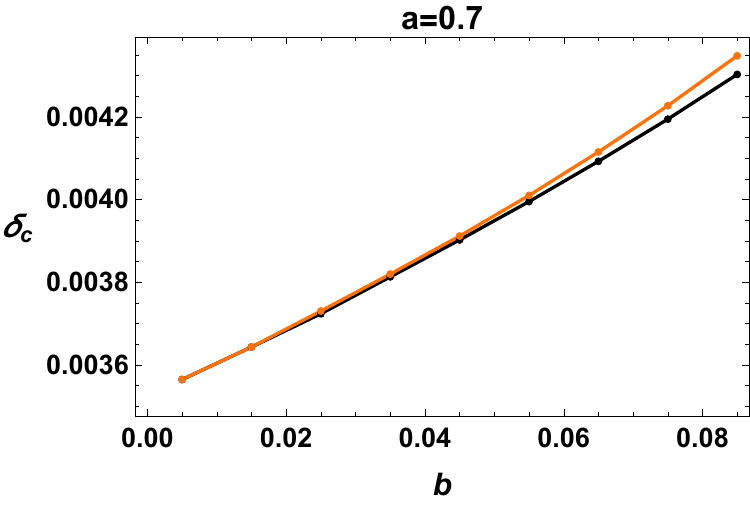} \>
			\includegraphics[scale=.59]{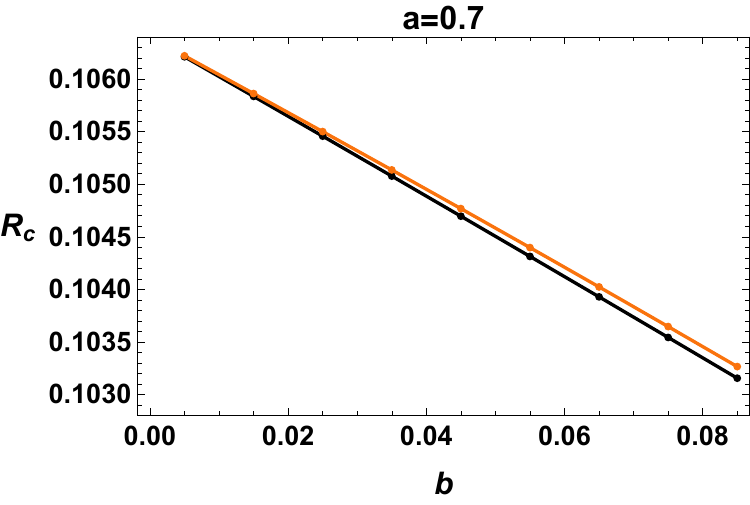} \\
			\includegraphics[scale=.59]{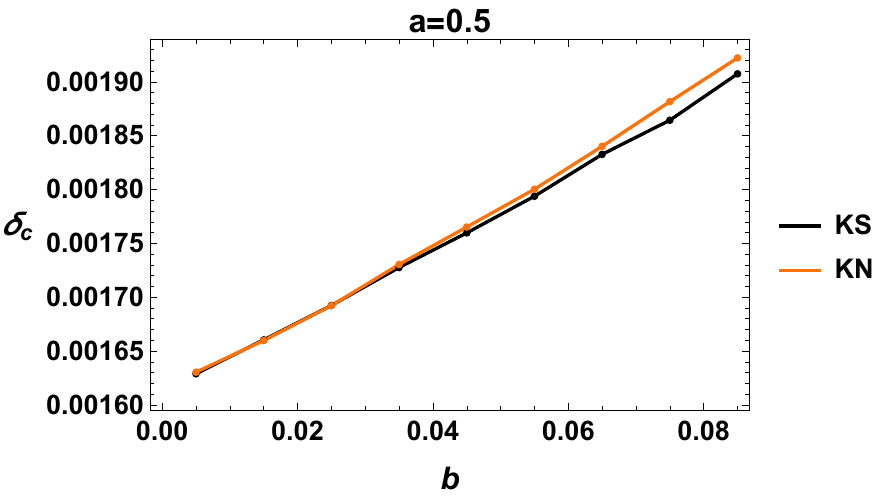} \>
			\includegraphics[scale=.59]{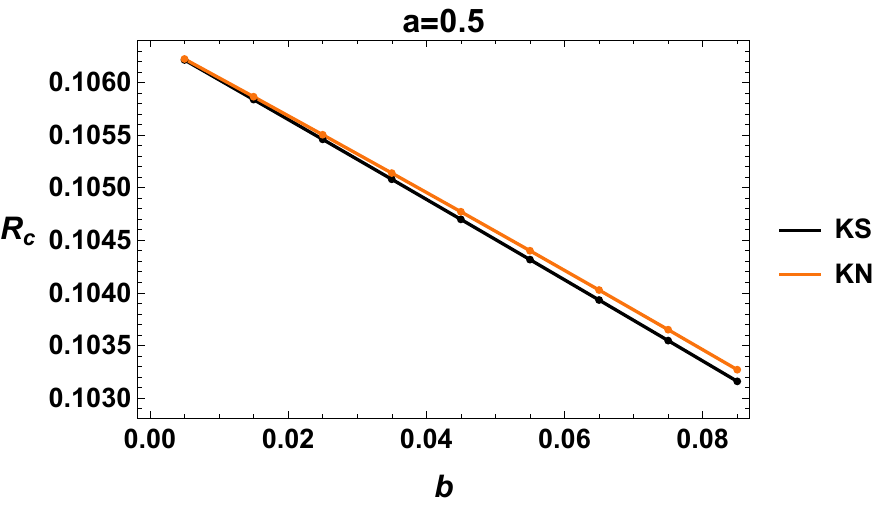} \\
			
				        \end{tabbing}
\caption{{\it \footnotesize {Astronomical observables  for different values of $b$ and  $a$ by taking  $\Lambda=-10^{-4}$ and  $m=1$. }}}
\label{Rb}
\end{figure}

To inspect  the geometric deformations of the black hole shadows, one usually  approach
  two parameters $R_c$ and $\delta_c$  providing   the size and  the shape approximations, respectively \cite{hioki2009measurement,amir2016shapes}.  Precisely, the size is characterized by three specific points being  top and bottom position of shadow ($x_t,y_t$), ($x_b,y_b$),  the point of reference circle ($\tilde{x}_p,0$). In this way,  the  point of distorted shadow circle ($x_p,0$) meets  the horizontal axis  at $x_p$.  Moreover, the distance between the two  letter points   is   controlled  by a parameter $D_c= \tilde{x}_p- {x}_p=2R_c-(x_r-x_p)$ \cite{amir2016shapes}.  $R_c$   is approximately  given by
\begin{equation}
\label{d1}
R_c=\frac{(x_t-x_r)^2+y_t^2}{2|x_t-x_r|}.
\end{equation}
However, the  distortion parameter  defined as  a ratio  of  $D_c$ and $R_c$  is given by
\begin{equation}
\label{ dddd}
\delta_c=\frac{|D_c|}{R_c}.
\end{equation}
To provide a deep comparative study concerning  the KN-AdS and the  KS-AdS  black holes, we analyse the astronomical parameters $R_c$ and $\delta_c$.   These two observables are plotted in  Fig.(\ref{Rb}) in terms of the $(a,b)$  reduced moduli space.

It has been observed that $R_c$, controlling the size,    decreases  by  increasing  the  parameters $b$,  $R_c$ is  almost  the same for both black holes even if we  vary the  parameter $a$.  For the value of $b$, between $0.005$ to $0.09$, $R_c^{KS}$ and $R_c^{KN}$  has almost same size for different values of  $a$. Concerning  the remaining  astronomical parameter $\delta_c$, controlling the distortion,  it   is plotted for both black holes in  right panels of Fig.(\ref{Rb}).  It follows  that for $a=0.2$,  the distortion parameter $\delta_c$ is almost zero for both types of black holes.  For  $a> 0.2$,  $\delta_c$  increases  by  increasing  the  parameter $a$. It has  been observed that  $\delta_c^{KN}$   is bigger  than $\delta_c^{KS}$  for values of $b$ above  $0.05$.  For values lower than $0.05$, however,   $\delta_c^{KN}$ and  $\delta_c^{KS}$ are equal. For  $b=0.005$,  the distortion  $\delta_c$ of the two black holes  coincides  even if we vary the parameter $a$. Otherwise, $\delta_c^{KN}$ is  bigger than  $\delta_c^{KS}$, showing that the distortion in  the KN-AdS spacetime is more  relevant  than the one  in  the  KS-AdS background.\\
  Having discussed the shadow shapes of the rotating and charged black holes with a negative cosmological constant, we move to  investigate  the  energy emission rate.

\subsection{Energy emission rate}
\begin{figure}[!htb]
		\begin{center}
		\centering
			\begin{tabbing}
			\centering
			\hspace{9.4cm}\=\kill
			\includegraphics[scale=.55]{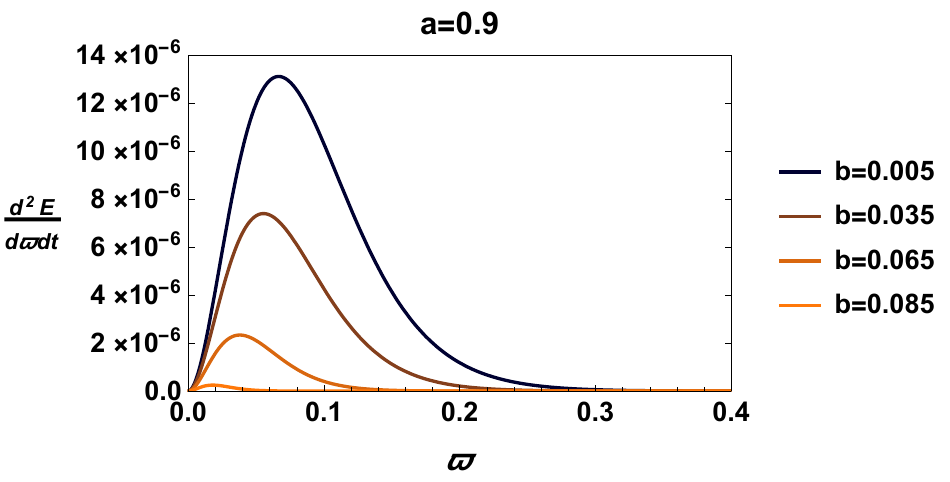} \>
			\includegraphics[scale=0.55]{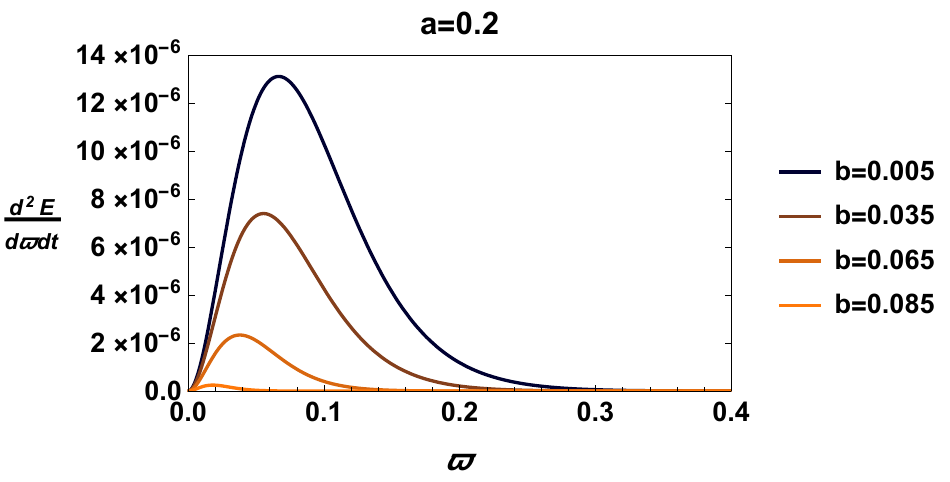} \\
		   \end{tabbing}
\caption{{\it \footnotesize {Energy Emission rate for KS-AdS black hole for different values of  $b$ and  $a$ by taking  $\Lambda=-10^{-4}$ and  $m=1$. } }}
\label{enmKS}
\end{center}
\end{figure}

\begin{figure}[!htb]
		\begin{center}
		\centering
			\begin{tabbing}
			\centering
			\hspace{9.4cm}\=\kill
			\includegraphics[scale=.55]{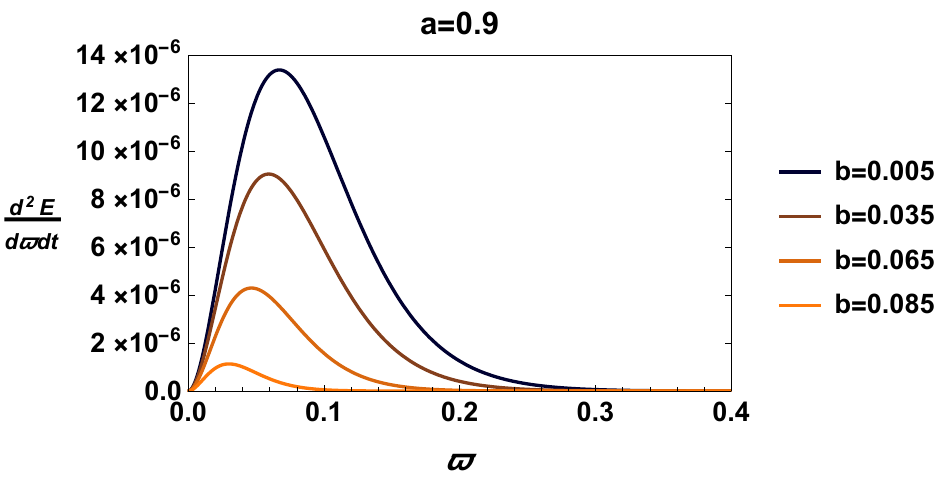} \>
			\includegraphics[scale=0.55]{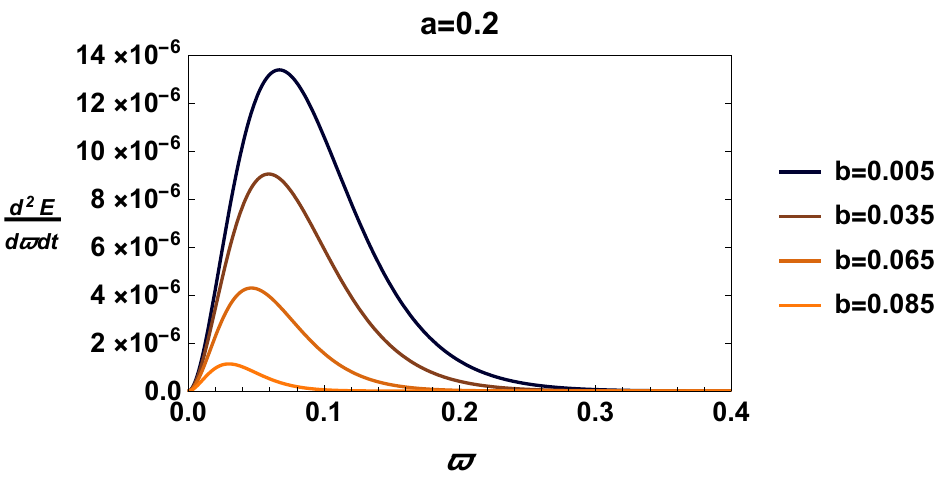} \\
		   \end{tabbing}
\caption{{\it \footnotesize {Energy emission rate of  KN-AdS black hole for different values of $b$ and  $a$ by taking  $\Lambda=-10^{-4}$ and  $m=1$.}}}
\label{enmKN}
\end{center}
\end{figure}

It has  been  known  that for,  a far distant observer, the  absorption cross-section approaches to the black hole shadow.  At very high energy, it is noted that   the  absorption cross-section  oscillates near to  a limiting constant value.  According to \cite{Wei:2013kza}, the  later being approximately equal to the area of the black hole shadow $(\sigma \sim \pi R_c^2)$  provides    the energy emission rate  expression  given by
\begin{equation}
\label{emirro}
\frac{d^2 E(\varpi)}{d\varpi dt}=\frac{2\pi^{3}( R^i_c)^{2}}{e^{\frac{\varpi}{T_{i}}}-1} \varpi^3, \qquad  i=KN, KS
\end{equation}
where $\varpi$ is the emission frequency.  In this relation, $T_{i}$ which denote  the temperature of the four-dimensional rotating and charged AdS black holes can be given   in terms of  the horizon radius $r_h^{i}$ $\left( \Delta_r (r^i_h) =0 \right)$.  It has been observed that  not all values of the temperature and the horizon radius are allowed for the the  rotating AdS  black hole  due  to the   presence of the parameter $a$ in the involved expressions. For the KS-AdS black hole, the temperature reads as
\begin{equation}
T_{KS}=\frac{1}{2 \pi  \left(a^2+(r_h^{KS})^{2} \right)} \left(r_h^{KS}+b- \frac{(b+r_h^{KS}) \left(a^2+4 b r^{KS}_h+2 (r_h^{KS})^2\right)\Lambda}{3}-m     \right).
\label{TKS}
\end{equation}
In Fig.(\ref{enmKS}), we plot the energy emission rate   as a function of the emission frequency $\varpi$   for  certain points of the $(a,b)$ moduli space.
 \\
It has been   remarked  from such a figure  that the effect of the twist parameter $b$ changes when we increase the rotation rate $a$. For $a=0.2$,  we concretely  observe that the energy emission rate increases with the decrease   in  the value of the  parameter $b$. For $a=0.9$, however,  the same  behavior is observed. Increasing  the  parameter $a$,  the emission rate remains constant.  \\
For KN-AdS  black hole,   certain  distinctions   appear.  Indeed,  the Hawking  temperature  is given by
\begin{equation}
T_{KN}=\frac{1}{2 \pi  \left(a^2+(r_h^{KN})^2\right)} \left(r_h^{KN}-\frac{r_h^{KN} \left(a^2+2 (r_h^{KN})^2\right)\Lambda}{3}-m \right).
\label{TKN}
\end{equation}
In Fig.(\ref{enmKN}),  these energetic aspects are  plotted  as a function of the emission frequency $\varpi$  for  different values of $a$ and  $b$.
 \\
For small values of the rotation rate $a$, it  has been seen   from  this  figure  that  the KN-AdS black hole  involves a  slower
 evaporation process  contrary to   the   KS-AdS  one. Increasing  the rotation rate parameter $a$,  however, we remark  that both black holes exhibit  similar behaviors. This behavior of  the  energy emission    can be clearly observed from the associated shadow radius and the temperature. This could be due to stringy  effects on the  black hole solutions. This  suggestion could be addressed in future works.

\section{Effects of  the cosmological constant on shadows}
\begin{figure}[th!]
		\begin{center}
		\centering
			\begin{tabbing}
			\centering
			\hspace{8.6cm}\=\kill
			\includegraphics[scale=.5]{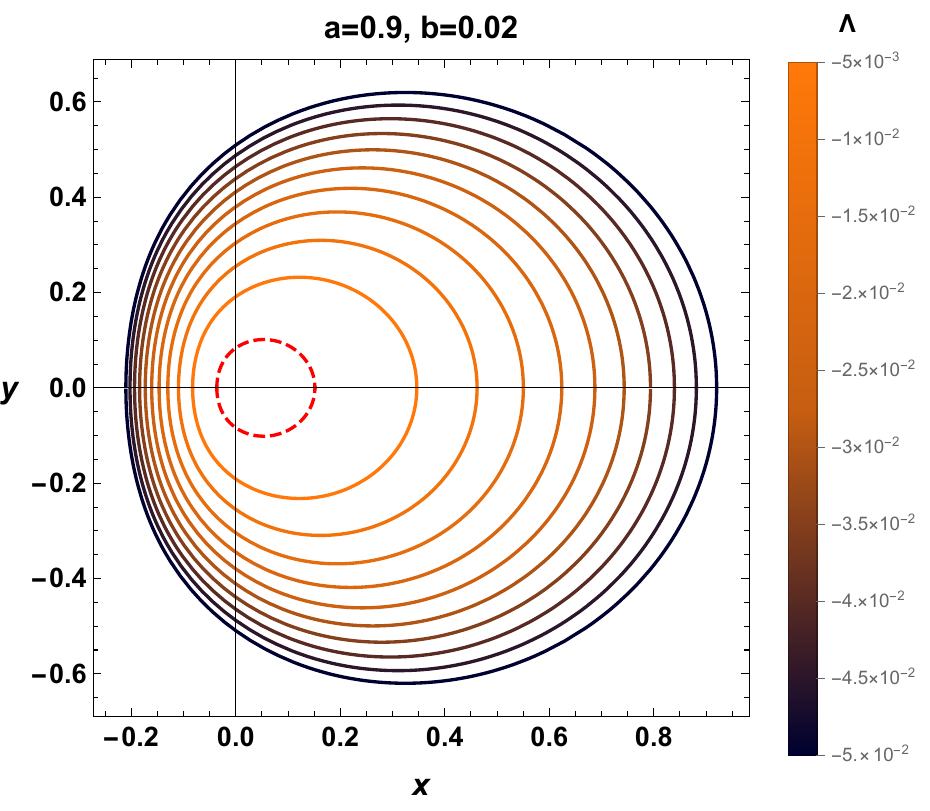} \>
			\includegraphics[scale=0.5]{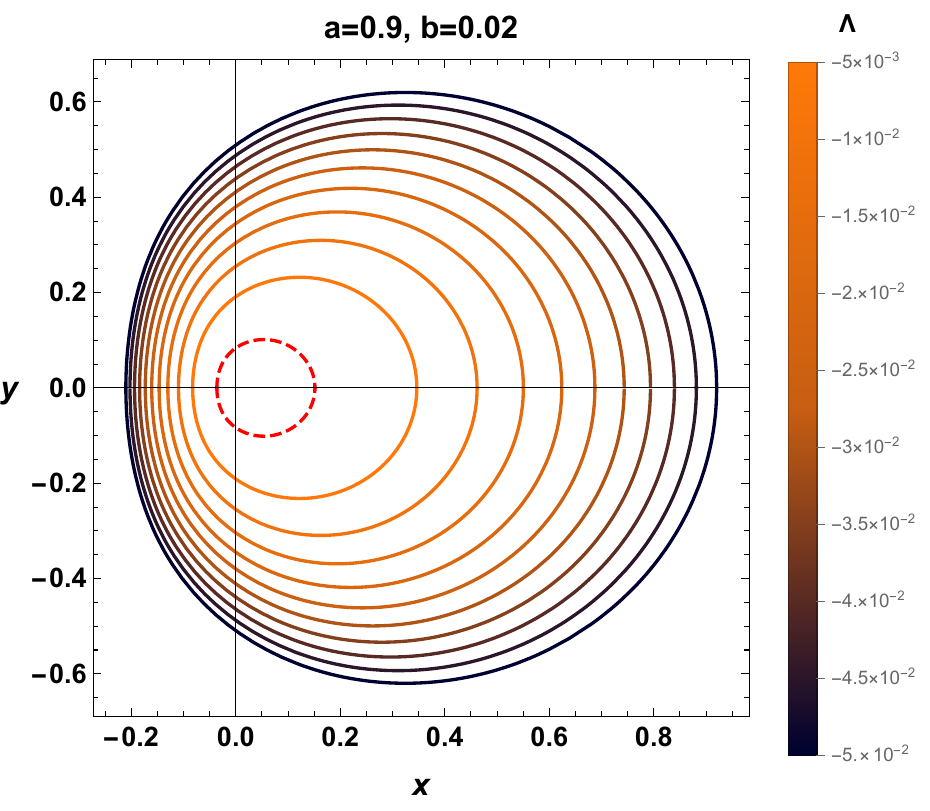} \\
		   \end{tabbing}
\caption{{\it \footnotesize Shadow behaviors  of  KN-AdS (left) and  KN-AdS (right)    for different values of the  cosmological constant $\Lambda$ for  $m=1$. In  all panels, the red and dashed lines  correspond to  $\Lambda=0$. The observer is positioned at $r_{ob}=50$ and $\theta_{ob}=\frac{\pi}{2}$}}
\label{Rb1}
\end{center}
\end{figure}

In this section, we  inspect the effects of the cosmological constant on shadows for  both classes of   charged  rotating  black holes.
In  Fig.(\ref{Rb1}), shadows with  negative   cosmological constant values  for KN-AdS and KS-AdS are compared  at   particular points  of the reduced  moduli space.

It follows from this figure that  the  same shadow size for  both  black holes has been observed.
Increasing  the value of the cosmological constant, the shadow size  decreases.  It has been  observed   that the cosmological constant affects the shadow size. Moreover, a similar comparative discussion has  been elaborated for zero cosmological constant  at  generic points of the moduli space associated with red circles. For small  values,  we  recover   the same result reported in  \cite{Carlos}.

\section{Conclusions and discussions}
More recently, it has been remarked that the shadows of  black holes has been considered as an active research  subject encouraged by the finding of   ETH international  collaborations.
Motivated  by such activities, we have  investigated the shadow of  charged rotating  black holes with a cosmological constant.    For  AdS  geometries, we have elaborated  two explicit models.
 First, we have  studied     shadow  optical behaviors   of    the KN-AdS  black holes.    Then, we have discussed  the naked singularity shadows  for a constrained region of the moduli space.  In order to unveil more data on such optical behaviors of  charged rotating  black holes with  a negative cosmological  constant,  we have  provided  a  comparative study.  Precisely, we have found
 that the KN-AdS  black hole possesses a small  shadow radius    compared to the KS-AdS one.   These optical  aspects    have been  consolidated by the energy emission rate and evaporation process. In  such an investigation,  relevant distinctions   have    appeared.  Precisely,  the KN-AdS energy emission rate  is small compared to the  KS-AdS  one   for some regions of the reduced  moduli space $(a,b)$. Moreover,  we have noticed  that certain $(a,b)$ regions  have different effects on  the KN-AdS black hole compared to  the KS-AdS one.\\
A close examination reveals that   the shadow of non-AdS KS and KN solutions have the same geometry. Moreover,  this also includes the size for certain values of the rotation and the  charge parameters \cite{Carlos}. This  result  has been recovered in the present study concerning the  KS-AdS and KN-AdS solutions. For certain values of $a$ and $b$, we  have obtained  the naked singularity geometry (an arc on the sky), which is the no domain of outer communication, of two black hole solutions in the presence of a cosmological constant. A similar  aspect  has been  observed for the Kerr–Newman–NUT black holes with a cosmological constant\cite{D1}.
This paper  comes up with many open questions. It would be of interest to inspect   behaviors associated with non  trivial  backgrounds  by considering either  the effects of the spacetime dimension  or external sources  provided by DE and DM.  It should be  also interesting also to  make contact with theoretical and  observational findings. We hope to address  elsewhere these open   questions.

\section*{Acknowledgment}
The authors would like to thank  H. Belmahi,   Y. Hassouni,  K. Masmar, M. B. Sedra, A. Segui,  and  E. Torrente-Lujan for discussions on related topics.
This work is partially supported by the ICTP through AF-13.

\end{document}